\documentclass[preprint]{ptephy_v1}%%%%%% to

\preprintnumber{XXXX-XXXX} %%% Insert preprint number here

\begin{document}
\title{$E(k,L)$ level statistics of classically integrable quantum systems based on the Berry-Robnik approach}
\author{\name{Hironori Makino}{1,\ast}, \name{Nariyuki Minami}{2,\dag}}
%%%%% Please use \thanks for contributed author details

%%%%%%%%%%% The \affil command should be used as \affil{Insert affiliation number here}{Insert author address here}
\address{\affil{1}{Department of Human and Information Science, Tokai University, 4-4-1 Kitakaname, Hiratsuka-shi, Kanagawa 259-1207, Japan}
\affil{2}{School of Medicine, Keio University, Hiyoshi 4-1-1, Kohoku-ku, Yokohama, Kanagawa 223-8521, Japan}
\email{makino@tokai-u.jp}
}

\begin{abstract}%
Theory of the quantal level statistics of classically integrable system, developed by Makino 
et al. in order to investigate the non-Poissonian behaviors of level-spacing 
distribution (LSD) and level-number variance (LNV)\cite{MT03,MMT09}, is successfully extended 
to the study of $E(K,L)$ function which constitutes a fundamental measure to determine most 
statistical observables of quantal levels in addition to LSD and LNV.  In the theory of Makino et al., the eigenenergy level is regarded 
as a superposition of infinitely many components whose formation is supported by the Berry-Robnik 
approach in the far semiclassical limit\cite{Robn1998}.  We derive the limiting $E(K,L)$ function in the 
limit of infinitely many components and elucidates its properties when energy levels show 
deviations from the Poisson statistics.
\end{abstract}

\subjectindex{A32}

\maketitle

%%%%%%%%%%%%%%%%%%%%%%%%%%%%%%%%%%%%%%%%%%%%%%%%%%%%%%%%
\section{Introduction}\label{sect1}
%%%%%%%%%%%%%%%%%%%%%%%%%%%%%%%%%%%%%%%%%%%%%%%%%%%%%%%%
One of the main objectives in the research field of quantum chaology 
is to elucidate the quantum manifestation of regular and chaotic features 
of classical dynamical systems\cite{Berry1989}.  Energy-level statistics, 
which were initially developed in nuclear physics\cite{Mehta1991}, 
played an important role in elucidating the universal properties of 
these manifestations.  In 1977, Berry and Tabor conjectured that, for 
a quantum system whose classical dynamical system is integrable, the 
energy eigenvalues in an unfolded scale\cite{Bohigas,MT03}, behave like uncorrelated 
random numbers from the Poisson process in the semiclassical limit, 
and that the fluctuation properties of these eigenvalues obey 
the Poisson statistics\cite{BT1977}.  This conjecture is in contrast 
with the conjecture of Bohigas, Giannoni, 
and Schmit of 1984, which states that the unfolded energy 
eigenvalues of a quantum system, whose classical dynamical system 
is fully chaotic, are well characterized by the GOE or GUE 
statistics of Random Matrix Theory(RMT) in the semiclassical limit\cite{BGS1984}.  
These two conjectures have been examined using various statistical 
observables, e.g., level spacing distribution(LSD), level number 
variance(LNV), spectral rigidity, mode fluctuation distribution, skewness and excess 
kurtosis, and these observables can be calculated from 
knowledge of the $E(K,L)$ function\cite{Mehta1991,Aurich1997,PR1999}.

The $E(K,L)$ function is defined as a distribution function that stands 
for the probability of finding $K$ levels in a randomly chosen energy-interval 
of length $L$.  For a given arbitrary value of non-negative 
integer $K=0,1,2\cdots$, $E(K,L)$ characterizes the fluctuation 
property of energy levels at the particular scale of 
$L=K$.  Once $E(K,L)$ is determined, the LSD is calculated 
as\cite{Mehta1991,Mehta1972}
\begin{equation}
P(K,L)=\frac{\partial^2}{\partial L^2}\sum_{j=0}^K (K-j+1)E(j,L).
\label{r1}
\end{equation}
The LSD $P(K,L)$ is introduced as a distribution function that 
denotes a probability density to find two adjacent levels of 
spacing $L$ containing $K$ levels in between.  The nearest-neighbor 
LSD(NNLSD), $P(0,L)$, which is frequently used to analyze the short-range 
spectral fluctuation, is a special case of $K=0$.  In a similar way, 
the LNV $\Sigma^2(L)$, skewness $\gamma_1(L)$ and excess 
kurtosis $\gamma_2(L)$, which are respectively, the two, 
three and four-point correlation functions, are calculated 
as $\Sigma^2(L)=C_2(L)$, $\gamma_1(L)=C_3(L)/{C_2(L)}^{3/2}$ and 
$\gamma_2(L)=C_4(L)/C_2(L)^2 -3$, respectively, where $C_n(L)$ is $n$th moment of 
the level number fluctuation around its average value $L$, obtained from $E(K,L)$ as 
\begin{equation}
C_n(L) = \sum_{K=0}^{+\infty}(K-L)^n E(K,L).
\label{(2)}
\end{equation}
Moreover, the spectral rigidity $\Delta_3(L)$, which is conventionally used to analyze 
the two-point correlation instead of $\Sigma^2(L)$, is also calculated from $E(K,L)$ as\cite{pandey1979}
\begin{equation}
\Delta_3(L) =\frac{2}{L^4}\int_0^L dS(L^3-2L^2 S +S^3)C_2(S).
\label{delta_3}
\end{equation}
In this way, it is quite important to determine the $E(K,L)$ function, which 
provides a basis for the energy level statistics.  For the Poissonian level 
sequence of the unfolded scale, $E(K,L)$ can be characterized by the Poisson distribution:
\begin{equation}
E_{\mbox{\tiny Poisson}}(K,L) = \frac{L^K}{K!}\exp{(-L)},
\label{Poisson_distribution}
\end{equation}
which obviously leads to the results from the Poisson 
statistics: $P(K,L)=L^K e^{-L}/K!$, $\Sigma^2(L)=L$, $\gamma_1(L)=L^{-1/2}$, 
$\gamma_2(L)=1/L$, and $\Delta_3(L)=L/15$.

Many works have examined the Berry-Tabor conjecture\cite{BT1977,Shnirelman,RV1998,MW1,MW2,MW3,MW4,MW5,MW6,CK1997,Biswas1991,MT03,MMT09,Richens1981,Shudo1989,BW1984,Aguiar2008}, 
and the statistical 
property of eigenenergy levels that the Poisson statistics can characterize 
is now widely accepted as a universal property of generic integrable quantum systems.  
However, the mechanism supporting this conjecture is still unclear, and deviation from the Poisson 
statistics is observed in some classically integrable systems that have a spatial or 
time-reversal symmetry.

One possible mechanism underlying the deviation from Poisson statistics has been 
proposed by Makino et al.\cite{MT03}, on the basis of the 
Berry-Robnik approach\cite{BR84,Robn1998,PR1999}.  We briefly review the 
outline as follows. For an integrable 
system, individual orbits are confined in each inherent torus whose surface is defined by holding its action 
variable constant, and the whole region of the phase space is densely covered with 
infinitely many invariant tori, which have infinitesimal volumes in the Liouville measure.  
Because of the suppression of quantum tunneling in the semiclassical limit $\hbar\to0$, 
the Wigner function of each quantal eigenstate is expected to be localized in the phase 
space region explored by a typical trajectory, and to form independent 
components\cite{Berry2083,Robn1998}.  For a classically integrable quantum system, 
the Wigner function localizes on the infinitesimal region in $\hbar\to0$
and tends to a $\delta$ function on a torus\cite{Berry237}.  Then, the eigenenergy levels 
can be represented as a statistically independent superposition of infinitely many 
components, each of which contributes infinitesimally to the level 
statistics. Therefore, if the individual spectral components are sparse enough, 
one would expect Poisson statistics to be observed as a result of the law of 
small numbers\cite{Fe1957}.  The statistical independence of spectral components 
is assumed to be justified by the principle of uniform semiclassical condensation 
of eigenstates in the phase space and by the lack of their mutual overlap, and thus 
can be expected only in the semiclassical limit\cite{Robn1998}.  This mechanism was 
initially introduced as a basis for the Berry-Robnik approach to investigate the 
energy level statistics of the generic mixed quantum system, and its validity is 
confirmed by numerical computations in the extremely deep semiclassical region 
which is called the Berry-Robnik regime\cite{PR1999REG}.

On the basis of this view, Makino and Tasaki investigated the NNLSD 
of systems with infinitely many components\cite{MT03}.  They derived the cumulative 
function of NNLSD, $M(L)=\int_0^L P(0,S)dS$, which is 
characterized by a single monotonically increasing function $\bar{\mu}(0,S)\in[0,1]$ of the nearest level spacing $S$ as
\begin{equation}
M(L)=1-\left[1-\bar{\mu}(0,L)\right] \exp{\left(-\int_0^L\left[ 1-\bar{\mu}(0,S)\right]dS \right)}. \label{(6)}
\end{equation}
The function $\bar{\mu}(0,S)$ classifies $M(L)$ into three cases: Case 1, Poisson distribution 
$M(L)=1-e^{-L}$ for all $L\geq0$ if $\bar{\mu}(0,+\infty)=0$; case 2, asymptotic Poisson distribution, 
which converges to the Poisson distribution for $L\to+\infty$, but possibly not for small spacing $L$ 
if $0<\bar{\mu}(0,+\infty)<1$; case 3, sub-Poisson distribution, which deviates from the Poisson 
distribution for $\forall L$ in such a way that $M(L)$ converges to 1 for $L\to+\infty$ more slowly 
than does the Poisson distribution if $\bar{\mu}(0,+\infty)=1$.  This argument is extended later to 
the study of LNV\cite{MMT09}, whose properties are evaluated for cases 1-3 as follows: 
Case 1, the LNV is the Poissonian $\Sigma^2(L) =L$; case 2, the LNV deviates from the Poissonian 
in such a way that the slope is greater than 1 for $L>0$ and approaches a number 
$\geq 1+2\bar{\mu}(0,+\infty)$ as $L\to+\infty$; case 3, the LNV deviates from the Poissonian 
in such a way that the slope is greater than 1 for $L>0$ and approaches a number $\geq3$ as 
$L\to+\infty$.  Therefore, the Berry-Robnik approach, when applied to classically integrable 
systems, allows the NNLSD and LND to deviate from the Poisson statistics.   

In this paper, extending the above arguments of Makino et al.\cite{MT03,MMT09}, we investigate 
the $E(K,L)$ function of a quantum system whose energy level consists of infinitely many 
independent components, and elucidate its property when the NNLSD of eigenenergy levels 
shows cases 2 and 3.  This paper suggests the possibility of a new statistical law to 
be observed in the $E(K,L)$ level statistics of classically integrable quantum 
systems.

The limiting $E(K,L)$ function is derived as follows: We consider a system whose phase space 
is decomposed into $N$ disjoint regions that give distinct spectral components.  The 
Liouville measures of these regions are denoted by $\rho_n(n=1,2,3,\cdots,N)$, which satisfy 
the normalization $\sum_{n=1}^N\rho_n=1$.  In the Berry-Robnik approach, these quantities 
are equivalent to the statistical weights of individual spectral components. 

When the entire sequence of energy levels is a product of statistically independent superpositions 
of $N$ subsequences, $E(K,L)$ is decomposed into the $E(K,L)$ function of subsequences, $e_n(k,L)$, as
\begin{equation}
E_N(K,L)=\sum_{\sum_{n=1}^N k_n = K}\prod_{n=1}^N e_n(k_n, L),\label{eq1-1}
\end{equation}
where $e_n$ satisfies the normalizations $\sum_{k=0}^{+\infty}e_n(k, L)=1$ and 
$\sum_{k=0}^{+\infty}k e_n(k, L)= L$.  In terms of the normalized level-spacing 
distribution $p_n(k,S)$ of the subsequence, $e_n(k,L)$ is described as
\begin{equation}
e_n(k,L) =\rho_n\int_L^{+\infty}dx \int_{x}^{+\infty} [p_n(k,S)-2p_n(k-1,S)+p_n(k-2,S)]dS,
\label{eq1-2}
\end{equation}
where $p_n(j<0,S)=0$, and $p_n$ satisties the normalization conditions 
$\int_0^{+\infty}p_n(k,S)dS=1$ and $\int_0^{+\infty}S p_n(k,S)dS=(k+1)/\rho_n$.  
Eq.(\ref{eq1-2}) is known as the formula in the theory of point process, which is 
derived as corollaries of the Palm-Khintchine theorem\cite{Minami}.

In addition to Eq.(\ref{eq1-1}), we introduce two assumptions that 
were introduced in Refs.\cite{MT03,MMT09}:\\
\indent
{\it{Assumption}} (i). The statistical weights of individual components vanish 
uniformly in the limit of infinitely many components: $\max_n \rho_n \rightarrow 0$ as $N\rightarrow +\infty$.\\
\indent
{\it{Assumption}} (ii). The weighted mean of the cumulative level-spacing distribution 
of spectral components, $\mu(k,S)\equiv \sum_{n=1}^N \rho_n\mu_n(k,S)$ 
with $\mu_n(k,L)=\int_0^L p_n(k,S)dS$, converges as $N\to +\infty$ to $\bar{\mu}(k,S)$, 
where the convergence is uniform on each closed interval: $S\in[0,L]$. It is noted that $\mu(k,S)$ is monotonically decreasing for increasing $k$.\\
In the Berry-Robnik approach, Eq.(\ref{eq1-1}) relates the level statistics 
in the semiclassical limit with the phase space geometry.

\indent
Under assumptions (i) and (ii), Eq.(\ref{eq1-1}) leads to the following new expression 
in the limit of $N\to +\infty$:
\begin{equation}
\bar{E}(K,L)=\alpha_K(L) e^{\beta(L)L}E_{\mbox{\tiny Poisson}}(K,L),
\label{3.1}
\end{equation}
where the factor $\alpha_K(L)$ and exponent $\beta(L)$ of the distribution function are described 
by the parameter function $\bar{\mu}(k,L)$.  When the 
lowest-order moment of this function shows $\bar{\mu}(0,L)=0$ for all $L$, one has $\alpha_K(L)=1$ for 
all $K$ and $\beta(L)=0$, and the limiting function $\bar{E}(K,L)$ of the whole energy sequence 
reduces to the Poisson distribution(\ref{Poisson_distribution}).  As shown in 
Ref.\cite{MT03}, this condition is expected to arise when the individual spectral components 
are sparse enough.  In general, one may expect $\bar{\mu}(0,L) > 0$, which corresponds to a 
certain accumulation of levels of individual components.  In this case, the limiting 
function $\bar{E}(K,L)$ deviates from the Poisson distribution.

The organization of this paper is as follows.  In Section \ref{sect2}, the limiting function 
$\bar{E}(K,L)$ is derived from Eq.(\ref{eq1-1}) and assumptions (i) and (ii).  In Section \ref{sect3}, 
the property of the limiting function $\bar{E}(K,L)$ is analyzed for cases 1--3, where the possibilities 
of deviation from the Poisson statistics are discussed.  In Section \ref{sect4}, the numerical 
investigation of the $E(K,L)$ function is carried out for the rectangular billiard, whose numerical 
results for the NNLSD have been shown to deviate from the Poisson distribution\cite{MT03,MMT09}.  
In Section \ref{sect5}, we discuss some relations between our results and those of related works.
%
%
%
%%%%%%%%%%%%%%%%%%%%%%%%%%%%%%%%%%%%%%%%%%%
\section{Limiting $\bar{E}(K,L)$ function}
\label{sect2}
%%%%%%%%%%%%%%%%%%%%%%%%%%%%%%%%%%%%%%%%%%%
%
Starting from Eq.(\ref{eq1-1}) and assumptions (i) and (ii), we derive the limiting function of 
$E_N(K,L)$ for a system with infinitely many components($N\to+\infty$).  First we transfer a 
sequence of nonnegative integers $\{k_n\}_{n=1,\cdots,N}$ of Eq.(\ref{eq1-1}) into a sequence of 
non-duplicate natural numbers $\{\kappa_m\}_{m=1,2,3,\cdots}$ with $\{d_m\}_{m=1,2,3,\cdots}$ 
being their individual duplications.  Then, the polynomial of Eq.(\ref{eq1-1}) is factorized 
using ratios $e_n'(\kappa_m,\rho_n L) \equiv e_n(\kappa_m, \rho_n L)/e_n(0,\rho_n L)$ as 
\begin{equation}
E_N(K,L)=E_N(0,L)\left[\sum_{M=1}^K \sum_{\sum_{m=1}^M d_m \kappa_m =K}\prod_{m=1}^M 
\frac{1}{d_m !} \left( \sum_{n=1}^N e_n'(\kappa_m,\rho_n L) \right)^{d_m} 
+\sum_{n=1}^N O\left(\rho_n^2\right) \right], 
\label{(2.1)}
\end{equation}
with
\begin{equation}
E_N(0,L) = \exp{\left(\sum_{n=1}^N \ln{e_n(0,\rho_n L)}\right)},
\end{equation}
where we have used properties 
$e_n(0,\rho_nL)^{-1}=1+O(\rho_n)$ and $e_n(k>0,\rho_nL)=O(\rho_n)$(see also Eq.(\ref{eq1-2})).  Since $e_n'(k>0,\rho_n L) =e_n(k,\rho_n L)+O(\rho_n^2)$ and Eq.(\ref{eq1-2}), $\sum_{n=1}^N e_n'(\kappa_m, \rho_nL)$ and $\sum_{n=1}^N \ln{e_n(0,\rho_n L)}$ are described by the weighted mean $\mu(k,S)=\sum_{n=1}^N \rho_n \int_0^S p_n(k,x)dx$ 
as
\begin{equation}
\sum_{n=1}^N e_n'(\kappa_m,\rho_n L) = \int_0^L dS \left[\mu(\kappa_m,S)-2
\mu(\kappa_m-1,S)+\mu(\kappa_m-2,S)\right] +\sum_{n=1}^N O\left(\rho_n^2\right),
\label{(2.3)}
\end{equation}
and
\begin{equation}
\sum_{n=1}^N \ln{e_n(0,\rho_n L)} = -L + \int_0^L \mu(0,S) dS +\sum_{n=1}^N O(\rho_n^2).
\label{(2.4)}
\end{equation}
Here, $\sum_{n=1}^N O\left(\rho_n^2\right)$ in the equations (\ref{(2.1)}), (\ref{(2.3)}) and (\ref{(2.4)}) shows the convergence,
\begin{equation}
\left| \sum_{n=1}^N O\left(\rho_n^2\right) \right| \leq C \max_n \rho_n \sum_{n=1}^N \rho_n = C \max_n\rho_n \to 0 \quad\mbox{as } N\to+\infty,
\label{(2.8)}
\end{equation}
which results from assumption (i).  Therefore, by applying assumption (ii), we have the 
limiting formula in the limit of $N\to+\infty$, 
\begin{equation}
\bar{E}(K,L) = \alpha_K(L) e^{\beta(L) L}  E_{\mbox{\tiny Poisson}}(K,L)
\label{(2.5)}, 
\end{equation}
where $\alpha_0(L)=1$,
\begin{equation}
\alpha_{K>0}(L)= \frac{K!}{L^K} \sum_{M=1}^K \sum_{\sum_{m=1}^M d_m \kappa_m =K}\prod_{m=1}^M 
\frac{1}{d_m !}\left(\int_{0}^{L} dS
\left[\begin{array}{c}\bar{\mu}(\kappa_m,S)\\-2\bar{\mu}(\kappa_m-1,S)\\+\bar{\mu}(\kappa_m-2,S)\end{array}\right]\right)^{d_m},
\label{(2.6)}
\end{equation}
and
\begin{equation}
\beta(L) = \frac{1}{L}\int_0^L  \bar{\mu}(0,S)dS.
\label{(2.7)}
\end{equation}
For $K=1-3$, the factors $\alpha_K(L)$ are specified as
\begin{equation}
\alpha_1(L)= \frac{1}{L}\int_0^L \left[1+\bar{\mu}(1,S)-2\bar{\mu}(0,S)\right]dS,
\label{(2.9)}
\end{equation}
\begin{eqnarray}
\alpha_2(L) &=&
\frac{2}{L^2}\int_0^L \left[\bar{\mu}(2,S)-2\bar{\mu} (1,S)+\bar{\mu}(0,S)\right]dS\nonumber\\
&& + \frac{1}{L^2}\left(   \int_0^L \left[1+\bar{\mu}(1,S)-2\bar{\mu}(0,S)\right]dS \right)^2,
\label{(2.10)}
\end{eqnarray}
and
\begin{eqnarray}
\alpha_3(L) &=&
\frac{6}{L^3}\int_0^L dS \left[\bar{\mu}(3,S)-2\bar{\mu}(2,S)+\bar{\mu}(1,S)\right]\nonumber\\
&&+ \frac{6}{L^3}\int_0^L dS \left[\bar{\mu}(2,S)
 -2\bar{\mu}(1,S)+\bar{\mu}(0,S)\right] \times\int_0^L dS \left[1+\bar{\mu}(1,S)-2\bar{\mu}(0,S)\right]\nonumber\\
&&+\frac{1}{L^3}\left(\int_0^L dS \left[1+\bar{\mu}(1,S)-2\bar{\mu}(0,S)\right] \right)^3.
\label{(2.11)}
\end{eqnarray}
%
%
%%%%%%%%%%%%%%%%%%%%%%%%%%%%%%%%%%%%%%%%%%%%%%%%%%%%%%%%%%%%%%%%%%%%%%%%%%%
\section{Properties of limiting  $\bar{E}(K,L)$ function}\label{sect3}
%%%%%%%%%%%%%%%%%%%%%%%%%%%%%%%%%%%%%%%%%%%%%%%%%%%%%%%%%%%%%%%%%%%%%%%%%%%
Since $\bar{\mu}(k,S)$ monotonically increases for $S\geq0$ and $0\leq \bar{\mu}(k,S)\leq1$, 
$\alpha_K(L)$ and $\beta(L)$ in the limit $L\to+\infty$ show
\begin{eqnarray}
\alpha_K(L)&=&K!\sum_{M=1}^K \sum_{\sum_{m=1}^M d_m \kappa_m =K} \frac{1}{L^{K-\sum_{m=1}^M d_m}}\nonumber\\
&& \times \prod_{m=1}^M 
\frac{1}{d_m !}\left(\frac{1}{L}\int_{0}^{L} dS
\left[\begin{array}{c}\bar{\mu}(\kappa_m,S)\\-2\bar{\mu}(\kappa_m-1,S)\\+\bar{\mu}(\kappa_m-2,S)\end{array}\right]\right)^{d_m} \\
&&  \longrightarrow\lim_{L\to+\infty}\left[ \frac{1}{L}\int_0^L (1+\bar{\mu}(1,S)-2\bar{\mu}(0,S))dS \right]^K \label{(2.12)}\\
&&  =\left[1+\bar{\mu}(1,+\infty)-2\bar{\mu}(0,+\infty)\right]^K\equiv \alpha_K(+\infty),\label{(2.13)}
\end{eqnarray}
and
\begin{equation}
\beta(L) \longrightarrow \bar{\mu}(0,+\infty).\label{(2.14)}
\end{equation}
Note that $K-\sum_{m=1}^M d_m = 0$ only when $M=1$ and $d_1=K$($\kappa_1=1$). 
From these convergences and the limiting value of the lowest-order function 
$\bar{\mu}(0,+\infty)$, all $\bar{E}(K,L)$ of $K=0,1,2\cdots$ are classified into the following three cases.

Case-1[$\bar{\mu}(0,+\infty)=0$]: $\bar{E}(K,L)$ is the Poisson distribution(\ref{Poisson_distribution}).  
This condition is equivalent to $\bar{\mu}(K,L)=0$ for all $L$ and $K$ since $\bar{\mu}(K,L)$ is 
monotonically increasing for $L$ and decreasing for $K$.  Thus, 
one has
\begin{equation}
\alpha_K(L)=1\mbox{ and }\beta(L)=0\mbox{ for all }L\mbox{ and }K.
\label{(2.15)}
\end{equation}

Case-2[$0<\bar{\mu}(0,+\infty)<1$]: $\bar{E}(K,L)$ deviates from the Poisson distribution, i.e., $\bar{E}(K,L)$ for a large value of 
$L$ is well approximated by the Poisson distribution, $\alpha_K(+\infty) L^K e^{-[1-\bar{\mu}(0,+\infty)]L} /K!$, where 
$\alpha_K(+\infty)$ is a value bounded by the inequality
$$
0\leq \alpha_K(+\infty) \leq \left[1-\bar{\mu}(0,+\infty)\right]^K.
$$
On the other hand for small value of $L$, $\bar{E}(K,L)$ may deviate from the 
Poisson distribution.  Since 
$\alpha_{K+1}(+\infty)=[ 1+\bar{\mu}(1,+\infty) -2\bar{\mu}(0,+\infty)]\alpha_K(+\infty)$ 
and $\bar{\mu}(0,+\infty) \geq \bar{\mu}(1,+\infty)$, the factor $\alpha_K(+\infty)$ monotonically 
decreases for $K$, i.e.,
\begin{equation}
\alpha_K(+\infty) > \alpha_{K+1}(+\infty) \mbox{ for all } K,
\label{(2.17)}
\end{equation}
where $\alpha_0(+\infty)=1$ and $\lim_{K\to+\infty}\alpha_K(+\infty)=0$. 

Case-3[$\bar{\mu}(0,+\infty)=1$]: $\bar{E}(K,L)$ in $L\to+\infty$ approaches 0 more slowly than 
does the Poisson distribution, where the factor corresponds to $\alpha_0(+\infty)=1$ and
\begin{equation} 
\alpha_{K}(+\infty)=0\\ \mbox{ for all }\\ K>0,
\label{(2.18)}
\end{equation}
and deviates from the Poisson distribution for all $L$.  

It should be noted that Case 1 and Case 3 are extreme cases where all factors 
$\alpha_{K}(L)$ of $K>0$ in the limit $L\to+\infty$ converge to 1 in Case 1 and to 0 in Case 3.

As is also shown in Ref.\cite{MT03}, one observes Case 1 if the scaled NNLSD of individual 
components $f_n(0,\rho_n S)=p_n(0,S)/\rho_n$, which satisfy 
$\int_0^{+\infty}f_n(0,x)dx=1$ and $\int_0^{+\infty}x f_n(0,x)dx= 1$ 
are uniformly bounded by a positive constant $D$ :  $\left| f_n(0,S)\right|\leq D$ ( $1\leq n\leq N$ ).  
Indeed, in $N\to+\infty$, the following holds:
\begin{equation}
\left| \mu(0,S)\right|\leq \sum_{n=1}^N \rho_n^2\int_0^S\left| 
f_n(0,\rho_n x)\right| dx 
\leq DS \sum_{n=1}^N \rho_n^2\leq DS\max_{n}\rho_n\sum_{n=1}^N\rho_n\to 0\equiv\bar{\mu}(0,S).
\label{(2.20)}
\end{equation}
Such a bounded condition is possible when the individual spectral components are sparse enough.

In general, one may observe Case 2 or Case 3, each of which corresponds to 
strong accumulation of energy levels, leading to a singular NNLSD of the 
individual components.  Such an accumulation can arise when the physical system 
has a symmetry\cite{BT1977,Shnirelman,RV1998,CK1997,Biswas1991,MT03,MMT09,Richens1981,Shudo1989,BW1984,Aguiar2008}.  
In the next section, we numerically analyze the $E(K,L)$ function for the 
rectangular billiard system whose NNLSD of the eigenenergy levels has been shown 
to obey Cases 2 and 3\cite{MMT09}.
%
%
%%%%%%%%%%%%%%%%%%%%%%%%%%%%%%%%%%%%%%%%%%%%%%%%%%%%%%%%%%%%%
\section{Numerical studies of rectangular billiard}
\label{sect4}
%%%%%%%%%%%%%%%%%%%%%%%%%%%%%%%%%%%%%%%%%%%%%%%%%%%%%%%%%%%%%
%
%
We analyze the property of $E(K,L)$ for a rectangular quantal billiard whose eigen-energy 
levels are given by $\epsilon_{n,m}=n^2+\gamma m^2$, where $n$ and $m$ are positive 
integers and $\gamma$ is the square ratio of two sides, $a$ and $b$, denoted as $\gamma=a^2/b^2$.  
The unfolding transformation $\{\epsilon_{m,n}\}\to\{\bar{\epsilon}_{m,n}\}$ is carried 
out by using the leading Weyl term of the integrated density of states, ${\cal N}(\epsilon)$, 
as $\bar{\epsilon}_{m,n}= {\cal N}(\epsilon_{m,n}) =\pi\epsilon_{m,n}/4\sqrt{\gamma}$.  
Berry and Tabor observed that the NNLSD of this system agrees with the Poisson distribution (Case 1) 
when $\gamma$ is far from rational, while it deviates from the Poisson distribution when $\gamma$ 
is rational\cite{BT1977}.  The deviation from the Poisson distribution was precisely analyzed 
for $\gamma=1$ by Connors and Keating\cite{CK1997}.  Working on the basis of Landau's 
number-theoretical result\cite{L1908}, they proved that the mean degeneracy of the eigen-energy 
levels increases logarithmically as the energy becomes higher.  This property has been confirmed 
numerically by Robnik and Veble in Ref.{\cite{RV1998}}, where the NNLSD $P(0,L)$ converges to the 
delta function in the high-energy(semiclassical) limit $\epsilon\to+\infty$.  In Ref.\cite{MMT09}, 
Makino et al. have shown that the logarithmic degeneracy of levels at $\gamma=1$ leads to 
$\bar{\mu}(0,+0)=1$ in the high-energy limit.  Since $\bar{\mu}(0,L)$ is monotonically increasing 
and $0\leq\bar{\mu}(0,L)\leq1$, the square billiard with $\gamma=1$ obviously 
shows $\bar{\mu}(0,+\infty)=1$, which corresponds to Case 3.

In this section, we evaluate numerically the behaviors of quantities that converge to 
$\alpha_K(L)$ and $\beta(L)$ in the semiclassical limit.  We carry out a 
numerical study for an irrational case in addition to a rational case ($\gamma=1$), which 
is described by a finite continued fraction of the golden mean number $(\sqrt{5}+1)/2$,
\begin{equation}
\gamma = 1+\frac{1}{1+}\frac{1}{1+}\cdots\frac{1}{1+}\frac{1}{1+\delta}
= [1;1,1,\cdots,1,1+\delta ],
\label{(3.2)}
\end{equation}
with an irrational truncation parameter $\delta\in [0,1)$.

Figure 1 shows semi-logarithmic plots of $E(K,L)$ for $K=0-4$ and $K=10$.  In each figure, we 
show results for three values of $\gamma$ corresponding to the (a)$41$st 
and (b) fourth approximations of the golden mean, and (c)$\gamma=1$.   
The solid curve in each figure represents 
the Poisson distribution (\ref{Poisson_distribution}).  Our analysis is 
valid for $K<< L_{\mbox{max}}$ as shown 
in Refs.\cite{RV1998,PR1999}, where $L_{\mbox{max}}$ is determined by the shortest 
period of classical periodic orbit\cite{Berry1985}, and it is calculated for the rectangular 
billiard as $L_{\mbox{max}}=\sqrt{\pi\bar{\epsilon}_{n,m}}\gamma^{-1/4}$.  We used eigen-energy 
levels $\bar{\epsilon}_{n,m}\in [ 100\times 10^{10},101\times 10^{10} ]$ 
corresponding to $L_{\mbox{max}}\sim 1.8\times 10^6$,  which is sufficiently large for 
our numerical study.  The numerical computation in this paper was carried out using the 
double-precision real number operation.  When the continued fraction is close to the 
golden mean number, $E(K,L)$ is well approximated by the Poisson distribution[plot (a)], 
and this result corresponds to Case 1 given in section \ref{sect3}.  In cases in 
which the continued fractions are far from the golden mean number, $E(K,L)$ clearly 
deviates from the Poisson distribution [plots (b) and (c)].  Since all levels at $\gamma=1$ are degenerate except those with $n=m$, $E(K,L)$ with odd $K$ is very small.

Figure 2 shows numerical plots of $\tilde{\alpha}_K(L)$ for $K=1-5$ and $K=10$, which are obtained by $E(K,L)$ as
\begin{equation}
\tilde{\alpha}_K(L)=\frac{K!}{L^K}\left(\frac{E(K,L)}{E(0,L)}\right).
\label{avava}
\end{equation}
In each figure, we show three results for $\gamma$ corresponding to plots (a)--(c) in figure 1.  
Note that function (\ref{avava}) is equivalent to $\alpha_K(L)$ in the 
semiclassical limit $\epsilon\to+\infty$.  In case $\gamma$ is close to 
the golden mean and $E(K,L)$ is well approximated by the Poisson distribution, 
$\tilde{\alpha}_K(L)$ agrees with $1$ very well [plot (a)].  On the other hand, in case $\gamma$ is far from the golden mean and 
$E(K,L)$ deviates from the Poisson distribution, $\tilde{\alpha}_K(L)$ approaches 
a number $\tilde{\alpha}_K(+\infty)$ such that $0<\tilde{\alpha}_K(+\infty)<1$[plot (b)], and this 
result corresponds to Case 2.  
In case $\gamma=1$, $\tilde{\alpha}_K(L)$ quickly converges to $0$ as $L\to+\infty$, and this 
result corresponds to Case 3.    It is quite interesting that $\tilde{\alpha}_K(L)$ of the 
4th approximation, whose result corresponds to Case 2, obviously shows relation (\ref{(2.17)}) of 
monotonically decreasing as $K$ increases as shown in Figure 3.

Figure 4 shows numerical plots of $\tilde{\beta}(L)\equiv 1+\frac{1}{L}\ln{E(0,L)}$ for the 
three values of $\gamma$ corresponding to plots (a)--(c) in Figure 1.  This 
function is equivalent to $\beta(L)$ in the semiclassical limit $\epsilon\to+\infty$, 
which satisfies $\beta(0)=\bar{\mu}(0,0)$ and $\lim_{L\to+\infty}\beta(L)=\bar{\mu}(0,+\infty)$.  
When $\gamma$ is close to the golden mean, $\tilde{\beta}(L)$ agrees with $0$ very well [plot (a)], 
and this result obeys the property of Case 1.  
On the other hand, in case $\gamma$ is far from the golden mean, $\tilde{\beta}(L)$ 
approaches a number $\tilde{\beta}(+\infty)$ such that 
$0<\tilde{\beta}(+\infty)<1$[plots (b)], and this result obeys the property of Case 2.  
However, in case $\gamma=1$ where Case 3 is expected to arise in the 
semiclassical limit, $\tilde{\beta}(L)$ does not reproduce $\beta(L)=1$ even 
in the region $L>>1$.  This is because we are not yet far enough in the high-energy 
region where $\tilde{\beta}(L)$ agrees well with $\beta(L)$.  
In order to estimate the convergence of $\tilde{\beta}(L)$ to 1, 
we analyze $\tilde{\beta}(+0)$, which can be described by the cumulative 
NNLSD $M(L)$ as $\tilde{\beta}(+0)=M(+0)$ (see Appendix \ref{append:A}).  
According to the theoretical prediction of Connors and Keating\cite{CK1997},
 and additional argument of Makino et. al.\cite{MMT09}, $M(+0)$ of the square billiard is described 
as $M(+0)\simeq 1-4 c/\pi\sqrt{\ln{\epsilon}}$, and thus we also have
\begin{equation}
\tilde{\beta}(+0) \simeq 1-\frac{4}{\pi} \frac{c}{\sqrt{\ln{\epsilon}}},\label{(3.4)}
\end{equation}
with $c\simeq 0.764$.  This indicates an extremely slow convergence of $\tilde{\beta}(+0)$ to 
1 as the energy $\epsilon$ becomes higher and $\beta(+0) \leq \beta(+\infty)= 1$ (Case 3) 
observed in the semiclassical limit ($\epsilon\to+\infty$).  
We finally confirm the approximate expression (\ref{(3.4)}).

Figure 5 shows $1-\tilde{\beta}(0)$ vs $4c/\pi\sqrt{\ln{\epsilon}}$ for various energy ranges.  The solid line represents 
the theoretical curve (\ref{(3.4)}), which is valid in the semiclassical (high energy) region.   Although we are not 
yet far enough in the high-energy region where $1-\tilde{\beta}(0)<<1$, the agreement between 
them is very good.
%
%
%%%%%%%%%%%%%%%%%%%%%%%%%%%%%%%%%%%%%%%%%%%%%%%%%%%%%
\section{Summary and Conclusion}\label{sect5}
%%%%%%%%%%%%%%%%%%%%%%%%%%%%%%%%%%%%%%%%%%%%%%%%%%%%%
%
%
%
%
The basic ideas of our study were to apply the Berry-Robnik approach to a classically integrable 
quantum system, whose phase space consists of infinitely many fine regions, and to discuss 
the possibility of deviations from the Poisson statistics.  In this paper, we successfully applied 
these ideas to the study of the $E(K,L)$ function which is one of the most fundamental 
observables in the research field of energy-level statistics.

In the Berry-Robnik approach, the quantal eigenfunctions, localizing on different 
phase-space regions in the neighborhood of the semiclassical(high-energy) limit, form 
mutually independent spectral components, where the statistical weight of each 
component corresponds to the volume ratio (Liouville measure) of the phase-space region.  
Therefore, we considered a situation where the system consists of infinitely many 
components and each of them contributes infinitesimally to the spectral statistics.  
Then, starting from the superposition formula (\ref{eq1-1}) and assumptions (i) and (ii), 
the limiting distribution function $\bar{E}(K,L)$ is derived, which is described by the monotonically 
increasing functions $\bar{\mu}(K,L),K=0,1,2,\cdots$ of the level-spacing $L$.  The limiting distribution 
function $\bar{E}(K,L)$ is distinguished into three cases: Case 1, Poissonian if $\bar{\mu}(0,+\infty)=0$; 
Case 2, Poissonian for large $L$, but possibly to be non-Poissonian for small $L$ if $0<\bar{\mu}(0,+\infty)<1$;
 and Case 3, non-Poissonian for all $L$ if $\bar{\mu}(0,+\infty)=1$.  Thus, we showed that deviations 
from Poisson statistics can be observed, not only in the properties of NNLSD and LNV as shown 
in the previous works of Refs.\cite{MT03,MMT09}, but also in the properties of individual $E(K,L)$ 
functions.  Note that cases 2 and 3 are possible when there is a strong accumulation of levels, 
which leads to a singular level spacing distribution of individual components, 
and such accumulation is expected to arise for a system that has a symmetry, e.g. 
spatial symmetry or time-reversal symmetry.

As was shown in the numerical studies of Refs.\cite{MT03,MMT09}, the rectangular billiard is 
one possible example by which to show case 2 in addition to case 1 when the aspect parameter of 
the system is irrational, and case 3 when the aspect parameter is rational.  In this paper, 
the numerical study of the rectangular billiard is extended to the analysis of the $E(K,L)$ function, 
where the theoretical arguments of $\bar{E}(K,L)$ for cases 1--3 are well reproduced.  
Similar results are expected also in the torus billiard\cite{Richens1981,RV1998}, equilateral-triangular 
billiard\cite{BW1984,Aguiar2008}, and integrable Morse oscillator\cite{Shudo1989}, where the deviations 
from Poisson statistics are reported to be associated with a spatial symmetry.

The limiting function $\bar{E}(K,L)$ obtained in the present paper, gives a basis on which to 
investigate the non-Poissonian behaviors of the other statistical observables.  For example, 
the $n$-point correlation function for a system with infinitely many components is 
calculated from the moment function,
\begin{equation}
\bar{C}_n = \sum_{K=0}^{+\infty}(K-L)^n \bar{E}(K,L),
\label{Cn}
\end{equation}
and this function for $n=2$ is described using $\bar{\mu}(K,L)$ in 
the following simple form\cite{MMT09}:
\begin{equation}
\bar{C}_2(L)=L+2\int_0^L \sum_{K=0}^{+\infty}\bar{\mu}(K,S)dS.
\end{equation}
For $0<\bar{\mu}(0,+\infty)\leq 1$, $C_2(L)$ provides the non-poissonian limiting LNV, 
$\bar{\Sigma}^2(L)=\bar{C}_2(L)$, whose slope is larger than that of the Poissonian 
LNV $\Sigma^2(L)=L$.  In a similar way, it is also possible to calculate 
the skewness $\bar{\gamma_1}(L)$ and excess $\bar{\gamma_2}(L)$ 
from $\bar{C}_3(L)$ and $\bar{C}_4(L)$ respectively, the LSD $P(K,L)$ of $K>0$ 
from Eq.(\ref{r1}), and $\Delta_3(L)$ from Eq.(\ref{delta_3}), which, for 
$\bar{\mu}(0,L)\not=0$, show the non-Poissonian behaviors.

This paper reveals the three different classes of $E(K,L)$ statistics possibly 
observed for the eigenenergy levels consisting of infinitely many components, 
and also suggests a possibility of new statistical laws (cases 2 and 3) to be observed 
in the classically integrable quantum systems that have spatial or time-reversal 
symmetry.  Further case study for individual physical systems will 
be shown elsewhere.
\begin{figure}[!h]
\centering\includegraphics[width=4in]{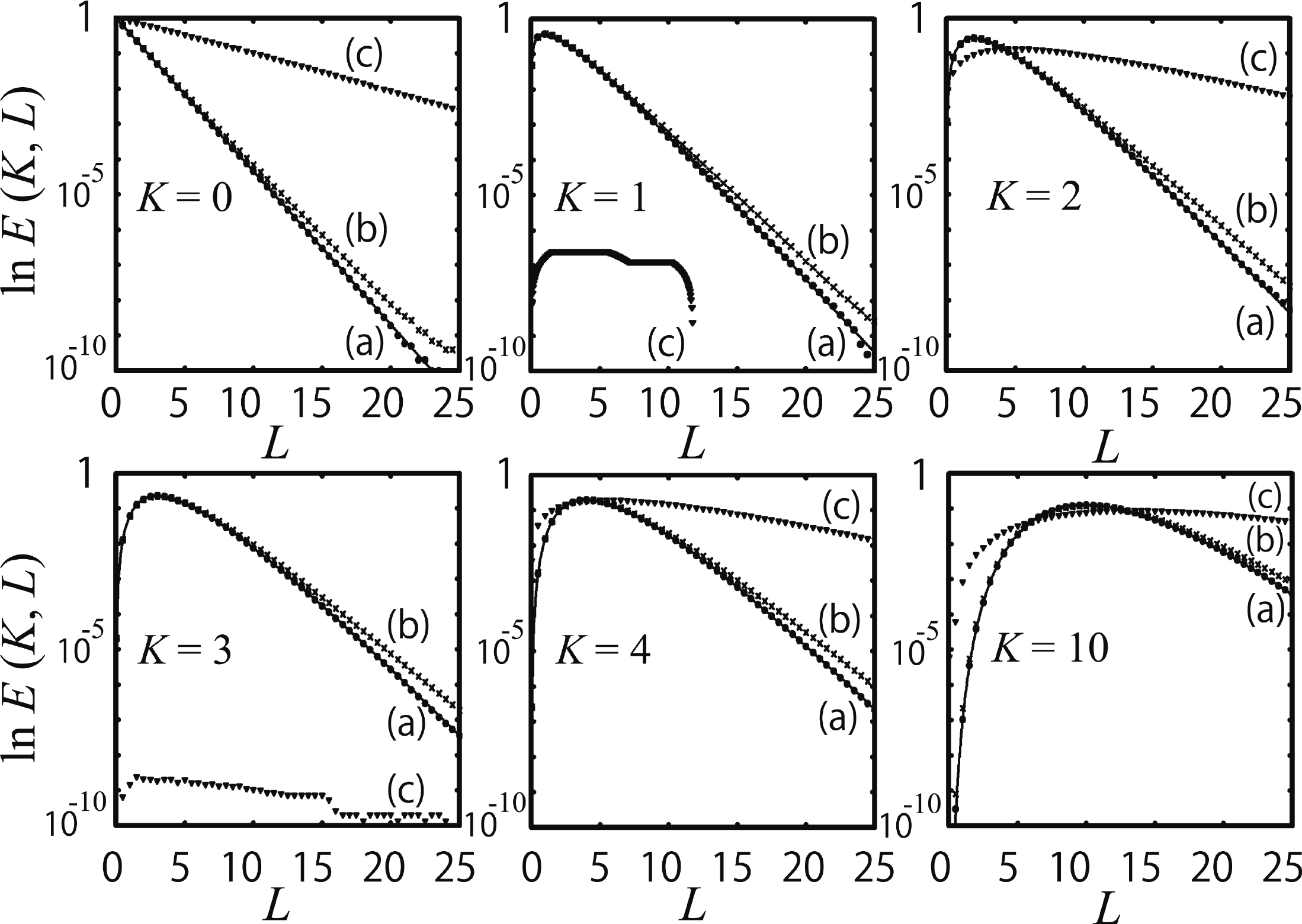}
\caption{Semilog plots of $E(K,L)$, $K=0-4$ and $10$ for the rectangular billiard 
systems[(a) 41th and (b) fourth approximations of $\gamma=(\sqrt{5}+1)/2$] and for the square billiard system[(c) $\gamma=1$].  
The truncation parameter is provided as $\delta=\pi\times 10^{-9}$. The solid curve 
corresponds to the Poisson distribution.}
\label{fig1} 
\end{figure}

\begin{figure}[!h]
\centering\includegraphics[width=5.5in]{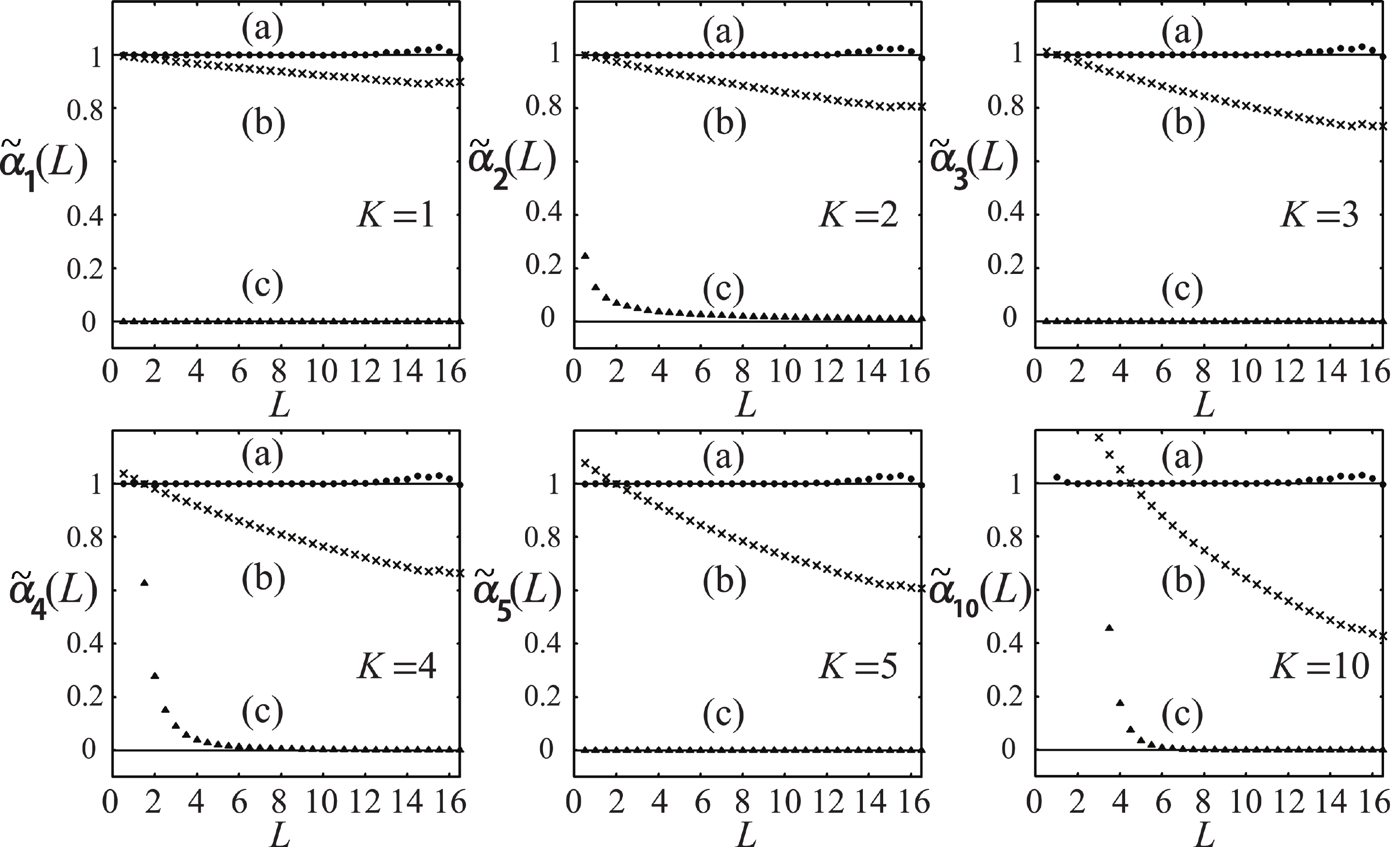}
\caption{
Numerical plots of $\tilde{\alpha}_K(L),K=1-4$ and $10$ for the rectangular billiard systems[(a) 41th and (b) fourth approximations of $\gamma=(\sqrt{5}+1)/2$] and for the square billiard system[(c) $\gamma=1$].  The solid line, $\tilde{\alpha}_K(L)=1$, corresponds to the Poisson distribution.}
\label{fig2}
\end{figure}

\begin{figure}[!h]
\centering\includegraphics[width=5.5in]{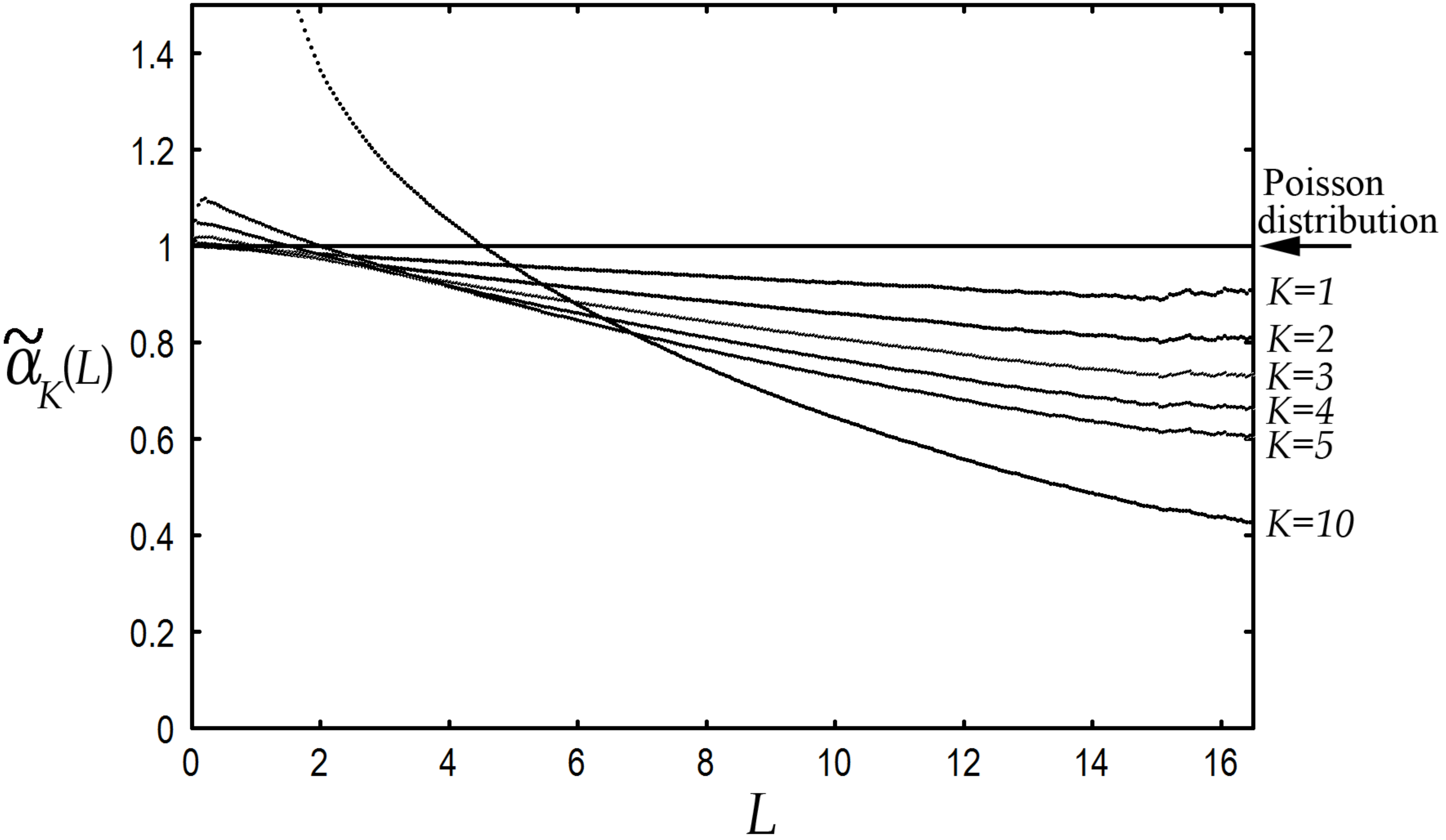}
\caption{
Numerical plots of $\tilde{\alpha}_K(L), K=1-5$ and $10$ for the rectangular billiard system with the 4th approximation of $\gamma=(\sqrt{5}+1)/2$.  The solid line, $\tilde{\alpha}_K(L)=1$, corresponds to the Poisson distribution.}
\label{fig3}
\end{figure}

\begin{figure}[!h]
\centering\includegraphics[width=3.5in]{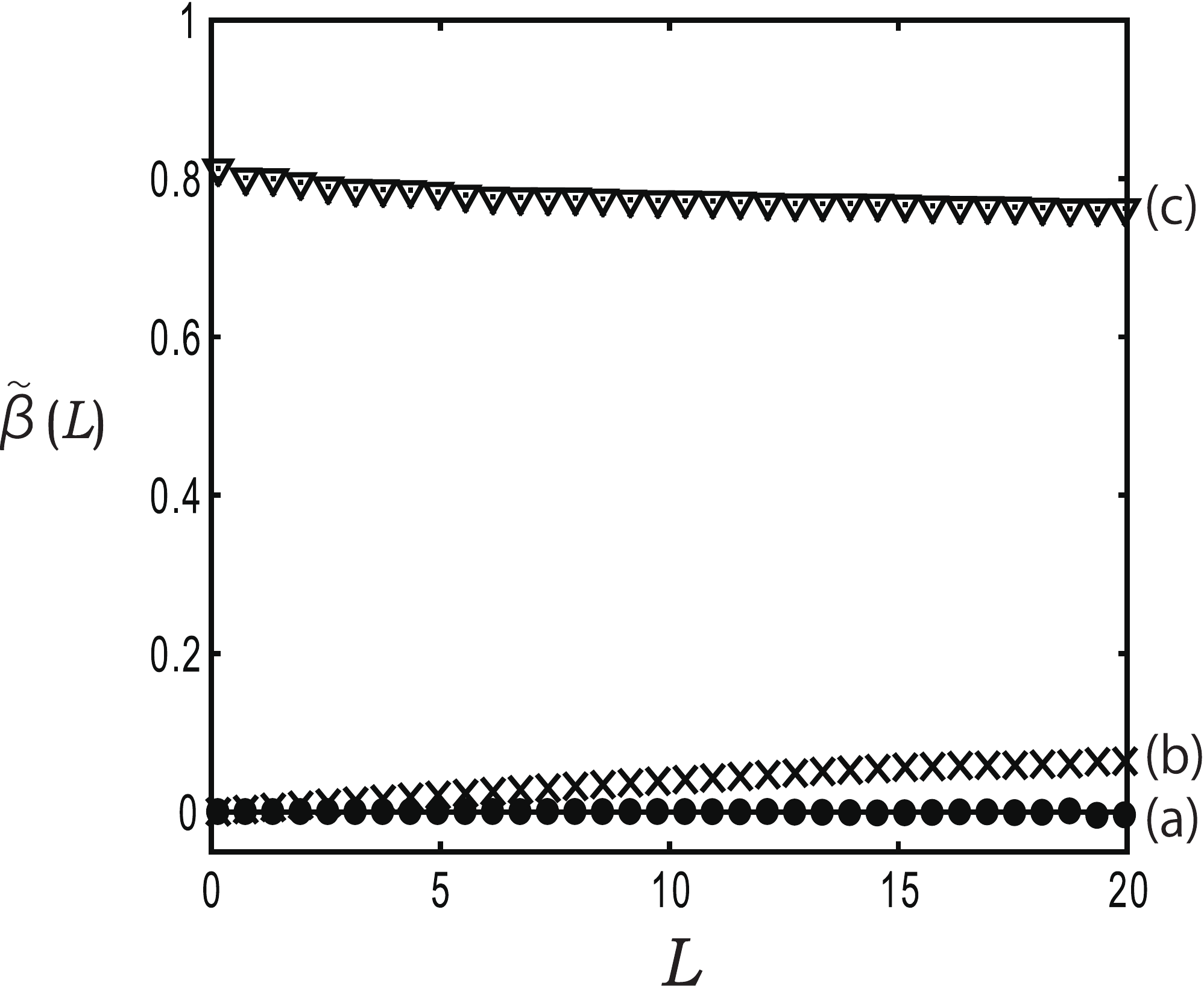}
\caption{
Numerical plots of $\tilde{\beta}(L)$ for the rectangular billiard systems[(a) 41th and (b) fourth approximations of $\gamma=(\sqrt{5}+1)/2$] and for the square billiard system[(c) $\gamma=1$].  The solid line, $\tilde{\beta}(L)=0$, corresponds to the Poisson distribution.}
\label{fig4}
\end{figure}

\begin{figure}[!h]
\centering\includegraphics[width=4in]{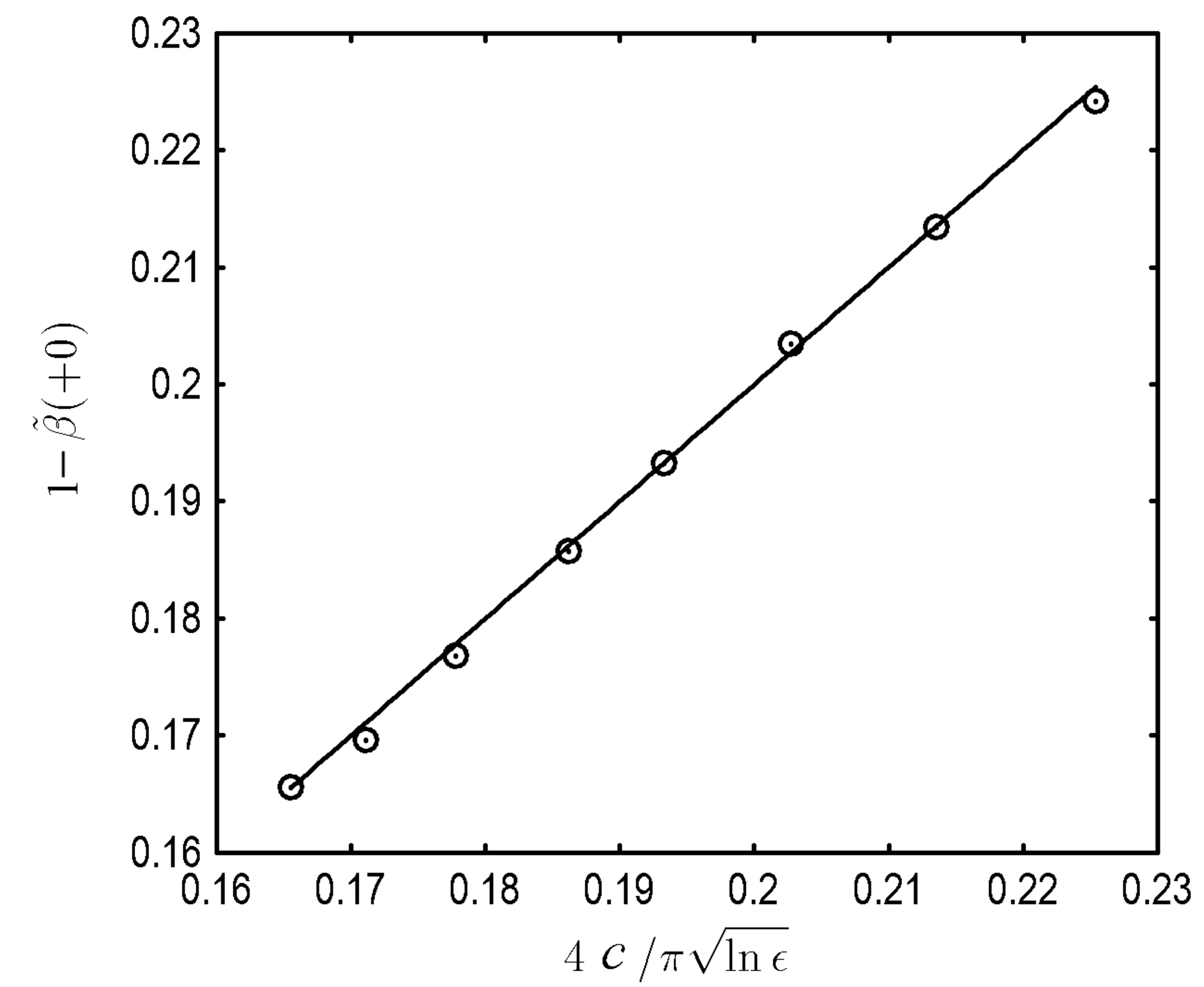}
\caption{
Numerical test of approximate expression (\ref{(3.4)}) for the square 
billiard system($\gamma=1$).  The solid line represents the theoretical 
prediction, witch is valid in the semiclassical limit $\epsilon\to+\infty$.  For each plots, we used $4\times 10^7$ unfolded energy levels.}
\label{fig5}
\end{figure}

\section*{Acknowledgment}
The authors would like to thank the late Prof. Shuichi Tasaki for penetrating comments and discussion which helped the authors to deepen and clarify their own arguments.

%  Insert the Acknowledgment text here.

% can use a bibliography generated by BibTeX as a .bbl file
% BibTeX documentation can be easily obtained at:
% http://www.ctan.org/tex-archive/biblio/bibtex/contrib/doc/

%\bibliographystyle{ptephy}
%\bibliography{sample}
%
% once the .bbl file has been generated then place the text in your article.

%\vfill\pagebreak

\appendix

\section{Appendix A: Derivation of equation $\tilde{\beta}(0)=M(0)$}
\label{append:A}
As corollaries of the Palm-Khintchine formula\cite{Minami}, 
the cumulative NNLSD $M(L)=\int_0^L P(S)dS$ is rewritten in terms of $E(0,L)$ as
\begin{equation}
M(S)=1+\frac{d}{dL}E(0,L) .
\label{A1}
\end{equation}
Since $E(0,0)=1$ and the equation (\ref{A1}), $\ln E(0,L)$ for $L<<1$ is expanded as
\begin{eqnarray}
\ln E(0,L)&=&L\frac{dE(0,L)}{dL}+O(L^2)\\
&=& -L\left[1-M(L)\right]+O(L^2),
\end{eqnarray}

and we have the relation

\begin{equation}
\tilde{\beta}(0)=1+\frac{1}{L}\ln E(0,L)|_{L=0}=M(0).
\end{equation}

\end{document}